**Female participation in science in the past 125 years: An analysis of the Matilda effect over time**

Felix Bittmann[#] & Lutz Bornmann*

#Leibniz-Institute for Educational Trajectories
Wilhelmsplatz 3
96047 Bamberg, Germany
E-Mail: felix.bittmann@lifbi.de
ORCID: 0000-0003-0802-5854

*Science Policy and Strategy Department
Administrative Headquarters of the Max Planck Society
Hofgartenstr. 8,
80539 Munich, Germany.
E-Mail: bornmann@gv.mpg.de
ORCID: 0000-0003-0810-7091

**Abstract**
The Matilda effect describes the systematic under-recognition and underrepresentation of women's contributions in science. While overall female participation in academic publishing has grown, the historical evolution of the Matilda effect across the science system remains less explored. In this study, we investigated this evolution over the past 125 years using the OpenAlex database, analyzing over 220 million publications from 1900 to 2025. We inferred the binary gender of approximately 60 million authors using a combination of deterministic approaches and large language models. To quantify the Matilda effect, we calculated the ratio between the overall share of female authors and the share of female corresponding authors. Because the corresponding authorship serves as a proxy for intellectual leadership and prestige, this ratio measures whether women are systematically excluded from receiving primary recognition for their collaborative work. Our analyses reveal a continuous increase in overall female participation, rising from approximately 10% in the 1910s to over 40% in recent years, interrupted only by temporary structural shocks during global wars. For most of the 20th century, the ratio between the two shares remained above 1, confirming a historical Matilda effect where women were systematically underrepresented in leadership roles relative to their overall participation. However, over the past two decades, this trend has globally inverted, indicating a decline of the traditional Matilda effect in academic publishing. Despite these overarching trends, significant disciplinary and regional differences persist. The physical sciences exhibit the lowest overall female share, while the health sciences approach parity. Regionally, Latin America closely approaches gender parity, whereas several Asian countries persistently lag behind and remain the only region where male authors continue to dominate corresponding authorship. Our findings provide comprehensive historical insights into the evolving integration of women in academic publishing and highlight remaining structural disparities.

**Keywords**
Matilda effect, Growth of science, Female participation, Bibliometrics

# 1 Introduction

Scientific discovery is a social enterprise governed by communication (Borgman 2007; Tahamtan and Bornmann 2022). The structural integrity of modern science relies on the codification, archiving, and public dissemination of research findings within international literature. The continuous production of text (documents) constitutes the collective, permanent record of human knowledge. From an anti-representationalist epistemological perspective, the systematic expansion of scientific understanding is functionally synonymous with the physical growth of scientific literature (Petrovich 2018). Scientific publications execute a fourfold institutional role: securing intellectual priority, facilitating broad dissemination, certifying quality via peer review, and establishing an archival recording for future reference and citation (Weingart 2025). Within the quantitative domain of the science of science, publications serve as the primary building blocks and analytical data sources (Milojević 2025). Their methodological utility stems from the standardized metadata they yield, including author names, institutional affiliations, citation trajectories, and keywords (Borgman and Furner 2002). These elements enable researchers to map the generation and diffusion of knowledge across various levels of aggregation, spanning from individual investigators to global institutional networks (Milojević 2025).

The empirical measurement of literature dynamics has a rich historical tradition (Evans, 2013). Derek de Solla Price pioneered this field in the 1960s and 1970s (de Bellis 2009), establishing that the volume of scientific literature scales exponentially over. Under the law of exponential growth, the rate of expansion at any given juncture is directly proportional to the total magnitude already achieved (Price 1986). This compounding acceleration imbues modern science with a property of 'immediacy', ensuring that the vast bulk of collective knowledge remains perpetually at the cutting edge (Wang and Barabási 2021). While early empirical efforts of measuring literature dynamics were frequently constrained by the fragmented coverage of historical literature databases, modern multi-disciplinary databases enable more accurate reconstructions of long-term scientific growth. For instance, bibliometric data from the Web of Science (WoS, Clarivate) spanning 2000 to 2008 indicates a 34% total expansion in article volume, with approximately 17% representing the real, intrinsic growth of science independent of database expanding policies (Michels and Schmoch 2012). Utilizing advanced statistical techniques such as segmented regression analysis on historical cited references, Bornmann and Mutz (2015) have identified distinct macro-historical growth phases. These models demonstrate a tripling of growth rates across successive historical eras: rising from less than 1% until the mid-18th century, accelerating to 2–3% up to the interwar period, and surging to 8–9% by 2012. By simultaneously integrating contemporary data streams from Dimensions (Digital Science), Microsoft Academic Graph (MAG), Scopus (Elsevier), and the WoS, most recent calculations place the unrestricted annual growth rate of modern science at 4.10%, corresponding to a doubling time of 17.3 years (Bornmann, Haunschild and Mutz 2021).

Although knowledge building in the (modern) science system should function as a universal meritocracy following the norms in the ethos of science (Merton 1973), the scientific enterprise displays structural stratification and inequality in research productivity and reward distribution (Allison 1980; Ziman 2000). Sociological studies demonstrate that this imbalance is driven by processes of cumulative advantage (Allison, Long and Krauze 1982; Zuckerman 1988). Over the course of scientific careers, minor initial differences compound to generate widening disparities in publications, resources, and recognition between the 'haves' and the 'have-nots' as cohorts age. While the disproportionate accumulation of rewards by prestigious scientists is known as the 'Matthew effect' (Merton 1968), Rossiter (1993) identified a critical, sex-linked counter-bias known as the 'Matilda effect'. Named after the 19th-century suffragette and critic Matilda Joslyn Gage – who documented how female innovations were routinely purloined or credited to men – the Matilda effect describes the systematic undervaluation, under-recognition, and historical erasure of women's scientific contributions, which are frequently misattributed to male colleagues (Rossiter 1993).

Empirical evidence tracking gender gaps in academic science reveals a complex and shifting landscape. Syntheses of literature from 2000 to 2020 show that while persistent claims of omnipresent sexism are countered by observed parity in tenure-track hiring, grant funding, journal acceptances, and recommendation letters, female scholars face ongoing disadvantages in teaching ratings and institutional salary distributions (Ceci, Kahn and Williams 2023). A foundational element of this disparity remains the gender productivity gap (Hess 1997). Across various cohorts, male researchers numerically dominate the most prolific productivity classes, resulting in an overall lower average publication volume for women (van den Besselaar and Sandstrom 2017). Bibliometric analyses of first-author representations across top journals demonstrate that while women achieve higher shares of first authorships in communication research and balance in sociology, male first authors heavily dominate political science and the most prestigious, top-tier publication outlets (Goyanes et al. 2025).

At the microscopic level, the Matilda effect may be sustained through biased perception and skewed co-authorship networks (Knobloch-Westerwick and Glynn 2013). Experimental testing demonstrates a cognitive bias among scholars: identical publication abstracts are rated as possessing significantly higher scientific quality, and evoke stronger collaboration interest, when they are assigned a male author name rather than a female one, particularly if the research topic is male-typed (Knobloch-Westerwick, Glynn and Huge 2013). Social network evaluations of co-authorship structures indicate that female scholars are frequently pushed to peripheral positions, with prominent research clusters remaining centered around male nodes (Srgado-Boj, Prieto-Gutierrez and Quevedo-Redondo 2021).

Classic historical precedents, such as the under-recognition of Barbara McClintock's genetic control systems by male molecular geneticists or the erasure of Zinaida Mulchenko from the authorship records of early Soviet scientometrics, are continuously replicated in the digital age (Kislenko and Kulczycki 2026; Wu 2024). Contemporary dissemination via social media platforms has failed to alleviate these inequities; instead, networks like X (formerly Twitter) amplify the Matilda effect, yielding higher relative citation gains and digital popularity for male scholars over their female peers (Song, Wang and Li 2024).

In this study, we aimed to bring together the two strands of the literature discussed above – the growth of science and the Matilda effect – and examined how women have participated in the historical development of science. While the Matthew effect states that scientists who are already famous receive disproportionately more recognition for the same achievements, the Matilda Effect describes the systematic suppression or failure to recognize the achievements of female scientists. We expected that despite increasing representation in the data, women appear less frequently in the reputable corresponding author position, as the Matilda effect proposes that their contributions are rendered 'invisible' or ranked lower in the hierarchy than the contributions of men. Our analysis made use of OpenAlex (Priem, Piwowar and Orr 2022), currently the most comprehensive database of scholarly metadata. Although Bornmann, Haunschild and Mutz (2021) had used several databases in their calculations of growth dynamics (without considering gender), they did not use OpenAlex. Since the research in this study analyzed the years between 1900 and 2025, it provides, to the best of our knowledge, the first quantitative insight into the Matilda effect in science from the early phases of modern science to the present. As we demonstrate above, referring to two case studies, this historical insight has previously been available only through qualitative studies. In addition to examining how the Matilda effect has changed over the past 125 years in general, we also investigated how much this effect varies by discipline and region.

# 2 Data and methods

## 2.1 Data and operationalization

We utilized with OpenAlex a fully open database launched by the nonprofit organization OurResearch on January 1, 2022, serving as a direct replacement for the discontinued MAG (Culbert et al. 2025; Priem, Piwowar and Orr 2022). Designed to adhere to the Principles of Open Scholarly Infrastructure (POSI), the platform provides completely unrestricted access to its complete dataset, application programming interfaces (APIs), and source code (Priem, Piwowar and Orr 2022). Since its inception, OpenAlex has rapidly established itself as a prominent alternative to commercial bibliographic databases (Donner 2025). This development is driven by its comprehensive literature coverage, continuous technological improvements, and the financial stability secured by OurResearch, which bolsters confidence among users regarding the database's long-term viability. In this study, we used a snapshot from April 2026 containing 492,361,307 entries from 109,695,206 authors, dating back several hundred years. We limited our observation period to 1900–2025, because bibliographic metadata regarding author names (which we used for gender classification) is generally less complete for historical publications. We included four types of publications in our study: journal articles, books, book-chapters, and preprints, according to the definition applied in OpenAlex. These types make up most of the publications and appear thus to be the most relevant types for science communication.

The major step of the study was to classify gender based on author names alone (since OpenAlex does not provide this data). It was our goal to infer the gender (either male or female) by the given name of an author. For this aim, we first cleaned the dataset thoroughly by removing entries without any names at all or without proper names that can be decoded (because the first names consisted only of initials). This reduced the number of names to about 94 million. We relied on tested deterministic approaches to infer the gender based on large available databases. These approaches classify gender information on common names, such as Lisa or Thomas, which makes inference secure and fast. We used the Python package *nomquamgender* (see https://github.com/ianvanbuskirk/nomquamgender) (Sánchez-Jiménez et al. 2024). *nomquamgender* was able to classify gender for 53.8 million names. We were left with a set of about 13.6 million names that were unclassified yet classifiable in principle: since the unclassified names were listed in full. For these cases, we utilized a large language model (LLM) to infer the gender from the complete name (full name of the author). We relied on a tested solution, the open-source model Nemo-Mistral (Herrero-Solana, González-Salmón and Robinson-Garcia 2026). Using this approach, we were able to classify the gender of about 6 million names. In total, we were able thus to assign a binary gender to a total of 59,890,585 names.

In total, we considered 223 million publications in this study published between 1900 and 2025.

## 2.2 Strategy of analysis

Our main goal was to analyze the share of female (corresponding) authors among publications and how this share has changed over the past 125 years.

First, we considered the share of female authors among all authors. For each publication, we counted the total number of authors of the paper, the total number of authors where a gender assignment was successful, and the number of authors that are female. We utilized the last two statistics to compute the share of female authors for a paper: number of authors that are female divided by number of authors where a gender assignment was successful.[1] We outline in Section 2.1 that we ignored authors that could not be assigned a gender. To receive results that are valid (based on our procedure), the assumption was that the successful assignment of gender is independent of gender itself.

Second, we considered not only the share of women among all authors but also the share of women among corresponding authors. In many scientific disciplines, an author's position in the byline serves as a reliable indicator of their contribution (Wilsdon et al. 2015; Wouters et al. 2015). Empirical evidence suggests that the first and last positions are the most significant in a scholarly publication (Wang and Barabási 2021). Typically, the first author undertakes the largest share of the work, from conducting experiments to drafting the manuscript, whereas the last author usually assumes a supervisory role and serves as the intellectual, financial, and organizational driving force behind the research. However, this contribution-based significance does not hold across all academic disciplines; in fields such as economics, author-listing protocols are frequently alphabetical, meaning that the physical order does not reflect relative individual contributions (Wilsdon et al. 2015; Wouters et al. 2015). Consequently, if a bibliometric analysis aims to identify and measure the primary investigator of a publication, the most appropriate method is to allocate credit to the corresponding author (de Moya-Anegón et al. 2013). The corresponding author is generally seen as the individual most responsible for the publication. Since a publication is primarily attributed to the corresponding author, this position is also prestigious and reputable among the authorship positions.

Utilizing the corresponding author position in this study minimizes measurement errors and reflects intellectual leadership in quantitative evaluations (Lee, Walsh and Wang 2015). The ratio between the share of women among all authors and the share of women among corresponding authors was used to estimate the Matilda effect over the last 125 years in science. A value greater than 1 indicates that the proportion of women among all authors is greater than the proportion of women among corresponding authors: Women are underrepresented in the prestigious position of corresponding author. A value greater than 1 thus suggests the presence of a Matilda effect in the assignment of author roles.

Third, we stratified the gender results on the primary topic domains, as defined by OpenAlex. These domains (disciplines) are health sciences, life sciences, physical sciences, and social sciences. We also grouped countries into groups to test for regional gender differences: Europe (excluding Russia and Turkey), Anglo-American countries (USA, Canada, Australia, and New Zealand), Asia (India, China, South Korea, and Japan), Africa, Americas (South and Middle America) and all other countries. The results for Asia are especially important because we can assume that gender inference is highly prone to error. The transliteration into the Latin alphabet permanently strips away the original logographic characters and their gender-specific semantic meanings. Furthermore, these transliterated names typically lack distinct phonetic markers, and varying naming conventions (such as placing the surname first) frequently cause algorithms to erroneously analyze the entirely gender-neutral family name instead. If we observe in our study a similar trend for Asia as for the other regions, this will indicate that the more difficult gender identification for Asian names did not affect the results. Our results indicate that this is not the case.

[1] We have tested whether weighting results by the share of classified gender makes any meaningful difference. As this was not the case, we report the unweighted results.

All analyses have been done using Python3 and Stata 18.0 (Bittmann 2019). The do-files to replicate all steps of the process are publicly available from BLINDED.[2] To ensure robustness, we have applied Multiple Imputation by Chained Equations (MICE) to our data. While MICE is a well-established method for handling missing data, its effectiveness depends on high-quality auxiliary variables capable of predicting missingness. In the OpenAlex dataset, however, potential auxiliary variables (such as the number of authors per publication, language, country of origin, domain, and publication type) proved to be weak predictors, typically explaining less than 5% of the variance. Despite the limited additional information MICE may provide in the specific context of this study, it remains a valuable tool for assessing the sensitivity of our results. When computing inference with MICE, the uncertainties around the point estimates are better considered, resulting in more credible inference as the share of missingness is directly integrated in the result. Hence, we argue that utilizing MICE is beneficial to achieve more reliable results. We minimized the risk of not over-interpreting the data: the data is volatile especially in earlier years, when case numbers were even lower.

# 3 Results

## 3.1 Matilda effect in academic publishing

Figure 1 illustrates the longitudinal evolution of the annual number of publications alongside the share of publications for which author gender information is available. Starting from approximately 50,000 records in 1900, the total publication volume exhibits a consistent pattern of exponential growth across the observation period. This long-term upward trajectory is temporarily disrupted by significant structural breaks corresponding to major global conflicts. Distinct contractions in overall publication output are visible following the onset of World War I, and again, more severely, during World War II. Following 1945, the publication volume resumes a rapid and steady expansion, reflecting the post-war institutionalization, globalization, and systematic funding of academic research. A final, distinct reduction in publication volume is visible in 2020, which can likely be attributed to the disruptive effects of the COVID-19 pandemic on global research output.

[2] We will provide these files after final publication.

In contrast to the overall publication volume, the share of publications with assignable gender metadata follows a markedly different trajectory. During the first half of the 20th century, the proportion of identifiable authors gradually declines, ultimately reaching a historical nadir immediately following 1945. It is only from the 1970s onward that the availability of author name information begins a sustained upward trend, eventually allowing for gender assignment in nearly 60% of publications in the most recent years. This accelerating coverage is likely driven by the advent of computational technologies and the internet (Krauss 2026). The digitalization of scholarly infrastructure facilitated the standardized storage, transfer, and accessibility of bibliometric data, making it increasingly possible to retrieve and utilize detailed author metadata.

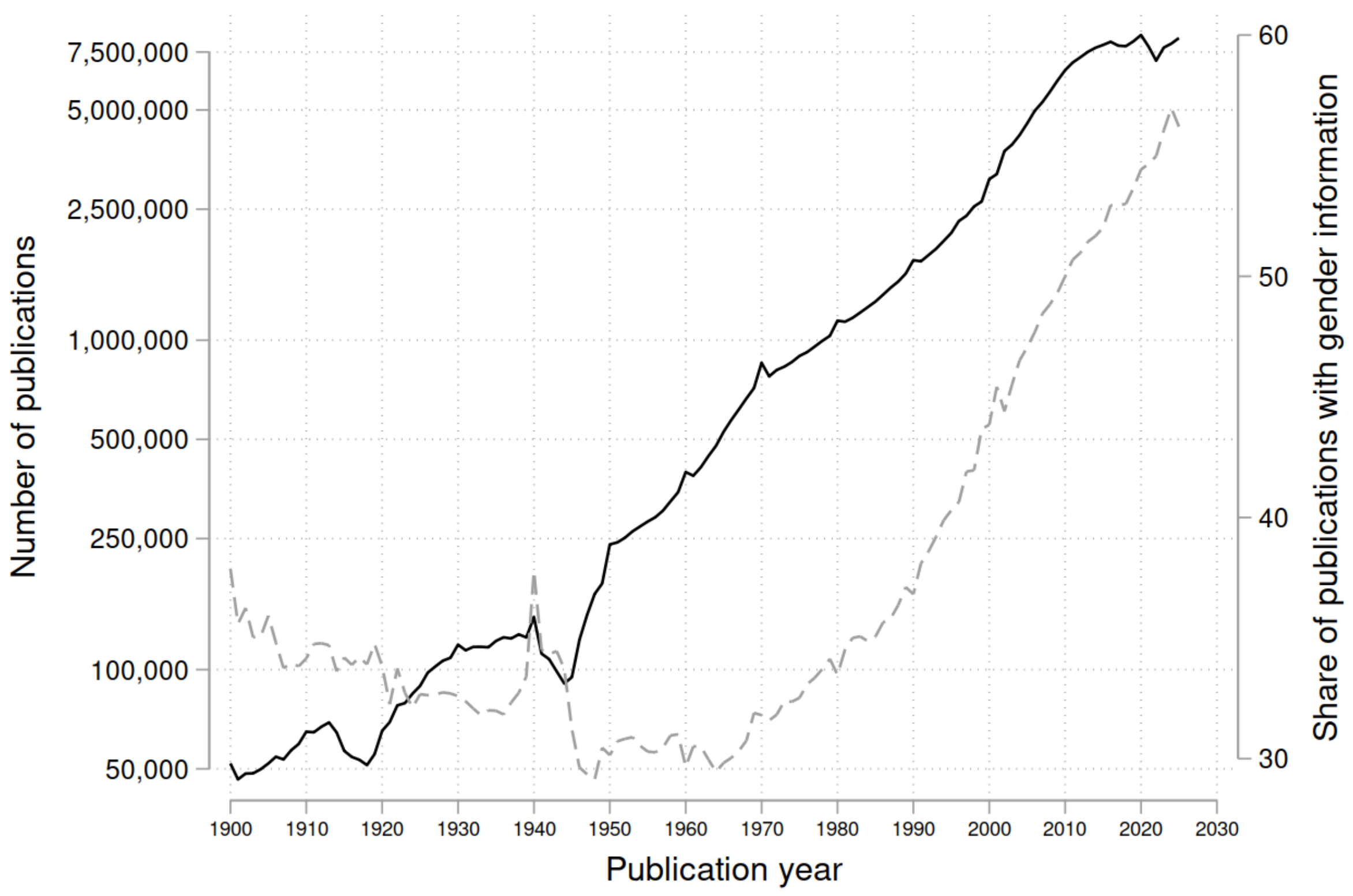

***Figure 1: Longitudinal growth of global publication volume and author gender metadata coverage.***
*Notes: The solid line displays the total number of annual publications (left y-axis). The dashed line represents the share of publications for which author gender could be assigned based on name information (right y-axis). Source: OpenAlex, own computations.*

Figure 2 integrates the gender dimension into the growth model by displaying the overall share of women among all authors alongside the ratio between the share of female authors and the share of female corresponding authors.

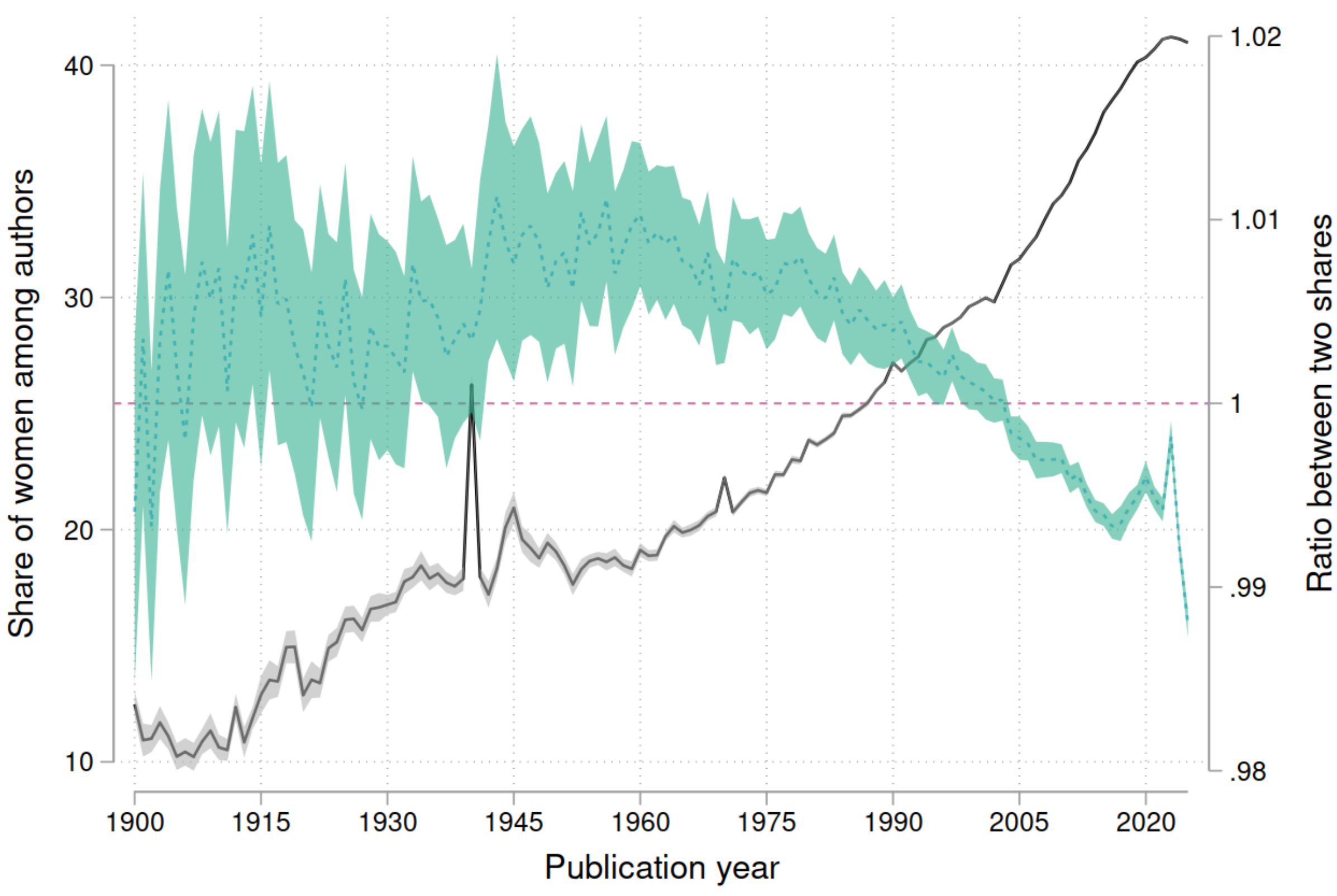

***Figure 2: Longitudinal integration of gender dimension and the Matilda effect in global publication growth.***
*Notes: The solid line displays the share of women among all authors (left y-axis). The dashed line represents the ratio between the share of female authors and the share of female corresponding authors (right y-axis), acting as an indicator for the Matilda effect. Source: OpenAlex, own computations. 95% confidence intervals included.*

Examining the overall share of female authors, distinct historical fluctuations become evident, most notably a temporary spike around 1940, alongside a similar trend during World War I, both followed by immediate reductions. With the escalation of World War II, a massive proportion of male researchers were drafted into military service or reassigned to classified, non-publishing defense projects, creating a structural vacuum in traditional civil research and university sectors. Women temporarily filled these vacated academic roles, leading to an artificial spike in their relative publication share. A second distinct peak in 1945 likely reflects the culmination of wartime research entering the publication pipeline just as the war ended. The sharp reduction immediately following was driven by demobilization, as returning male veterans reclaimed their academic positions and effectively displaced the women who had temporarily held them. The years leading up to 1970, marked by a third distinct peak, were characterized by the momentum of second-wave feminism, civil rights movements, and a massive structural expansion of the higher education sector across Western countries. This expansion created an unprecedented wave of hiring and provided a window of opportunity for female academics to enter the system.

When considering the ratio between the share of all female authors and corresponding female authors, the data reveals a systemic structural disadvantage for women in leadership roles for most of the observation period. Up until the year 2004, the data indicates that even as the overall share of female scientists working on publications rose, the lead role was usually taken by a man, confirming a historical Matilda effect in scientific authorship. However, over the past 20 years, this distribution has reverted as the ratio fell below the parity limit. This shift signifies that the traditional Matilda effect has not been observed in the last two decades; on the contrary, women are now slightly over-represented when it comes to taking the lead on a publication, suggesting an emergent advantage for women over men in this regard.

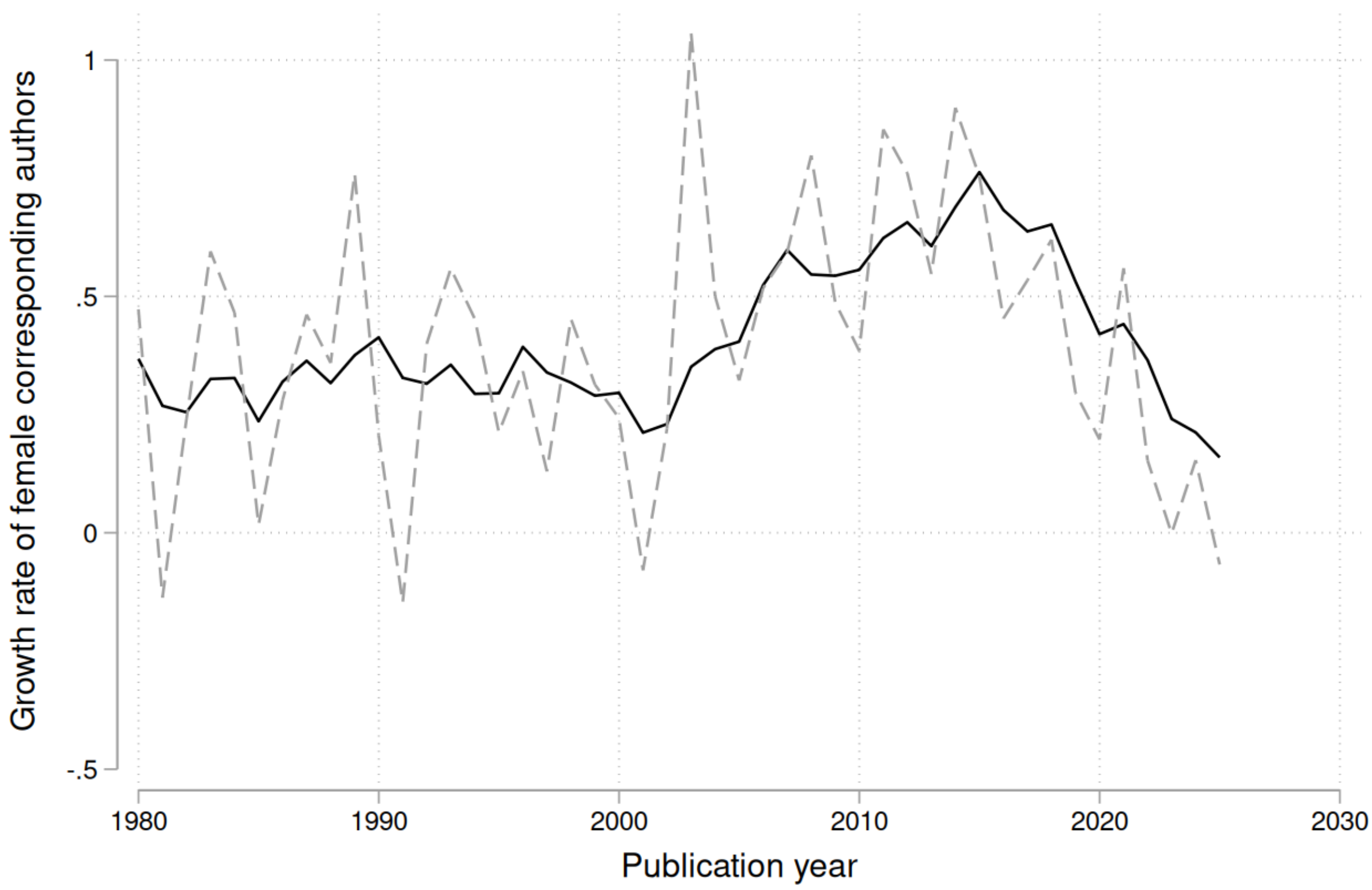

***Figure 3: Annual growth rate of female corresponding authorship (1980–2025).***
*Notes: The dashed line represents the raw annual growth rate (derivative) of the share of women among corresponding authors. The solid line displays a smoothed version of the data to visualize the underlying trend. Source: OpenAlex, own computations.*

Over the entire observation period from 1900 to 2025, the average annual increase in the share of female corresponding authors amounts to approximately 1.17%. When restricting the analysis to the period following major historical and structural breaks from 1975 onward, this average annual growth rises slightly to 1.28%. To provide a more granular understanding of these temporal dynamics, Figure 3 plots the derivative, or the annual growth rate, of the female corresponding authorship share from 1980 onward, displaying both raw values and a smoothed trend line for clearer visualization. The data demonstrates that while there has been a continuous increase in female representation among corresponding authors throughout this period, the momentum of this growth has fluctuated significantly. Between 1980 and the early 2000s, the growth rate remained approximately constant, hovering around 0.4 percentage points per year. This plateau was followed by a distinct acceleration phase spanning approximately 15 years, during which the annual growth rate surged to a peak of roughly 0.7 percentage points. However, the data reveals a significant deceleration following this peak. In the most recent years of the observation period, the growth rate has slowed down considerably, dropping to levels of about 0.2 percentage points. Ultimately, this indicates that after a sustained, 15-year period of robust acceleration in the inclusion of women as corresponding authors, the momentum is currently leveling off, returning to rates comparable to – and even slightly below – those observed prior to the rapid growth phase.

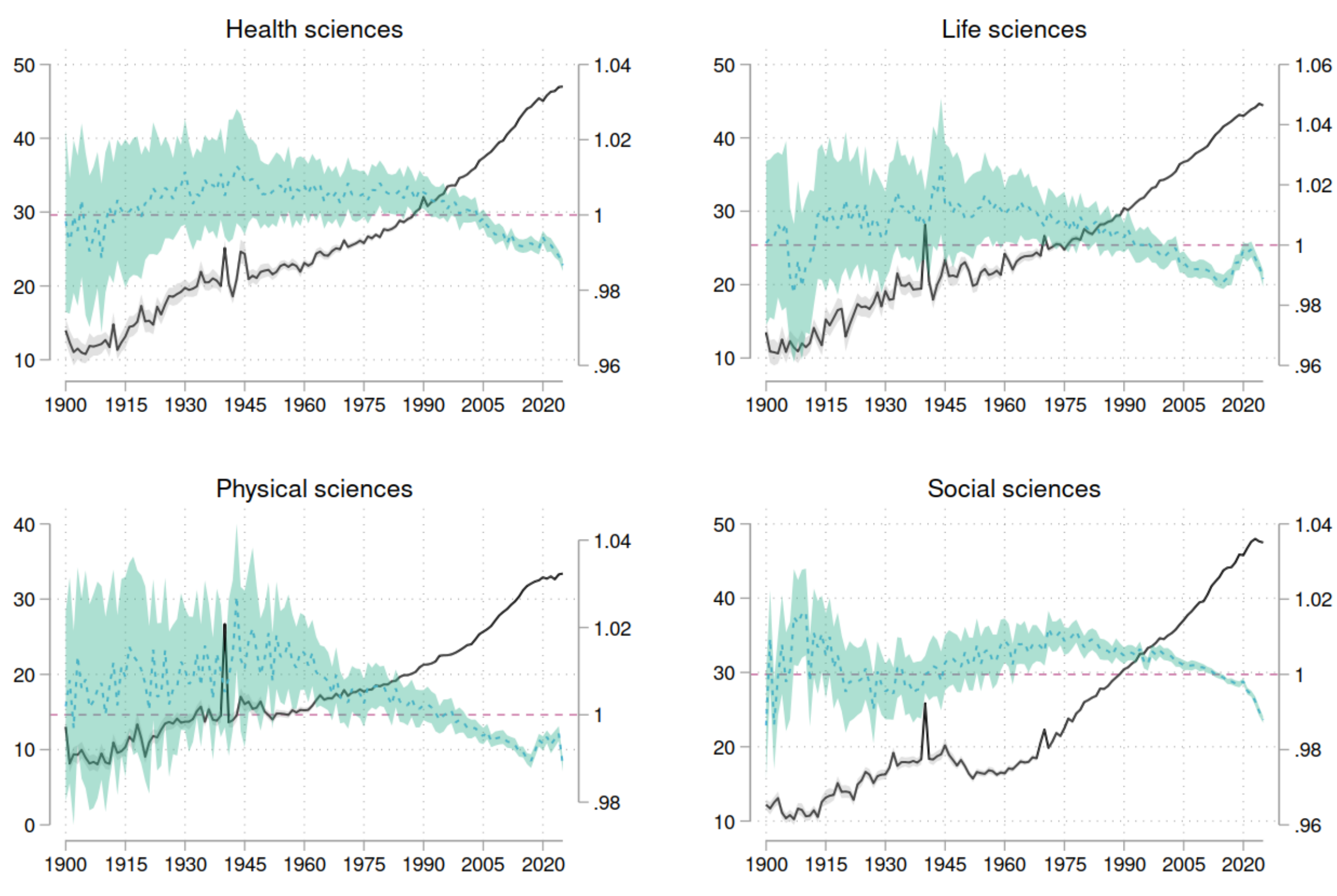

***Figure 4: Longitudinal trends in female authorship and the Matilda effect by discipline.***
*Notes: The solid line shows the overall share of female authors (left y-axis). The dashed line displays the ratio between the share of female authors and female corresponding authors (right y-axis), serving as an indicator for the Matilda effect. Source: OpenAlex, own computations. 95% confidence intervals included.*

Figure 4 illustrates the two metrics from Figure 2 across four major scientific disciplines (health, life, physical, and social sciences) from 1900 to 2025: the overall share of female authors and the ratio of the female authorship share to the female corresponding authorship share. A ratio greater than 1 indicates the presence of a disciplinary Matilda effect, where women are structurally underrepresented in the senior, leadership role of corresponding author relative to their overall presence in the discipline. A universal, long-term upward trajectory in overall female representation is evident across all disciplines. Starting from a baseline of approximately 10% in the early 20th century, female authorship experiences substantial growth, accelerating significantly from the 1970s onward. Like broader global authorship trends, these disciplinary trajectories exhibit distinct historical shocks, most notably a sharp, temporary spike around 1940 across all disciplines, indicative of the structural displacement of male researchers during World War II and a brief period of female mobilization in academic publishing.

When examining the ratio between two shares, a consistent historical pattern emerges across the entire scientific spectrum. For most of the 20th century, the ratio remained persistently above 1 in all four disciplines, confirming a historical, systemic Matilda effect. Even as women increasingly entered scientific disciplines and contributed as authors, they were systematically underrepresented as corresponding authors. However, a fundamental structural shift is visible over the past two to three decades, as the ratio has steadily declined and crossed the parity threshold in all disciplines. Currently, the ratio in all disciplines sits slightly below 1, indicating that the traditional Matilda effect regarding corresponding authorship has on average declined and inverted over the past 20 years. In recent years, women are slightly overrepresented as corresponding authors relative to their overall baseline authorship share.

Despite these shared overarching patterns, a persistent hierarchy remains regarding overall female inclusion, and the specific temporal trajectories of the Matilda effect reveal distinct disciplinary nuances. The health sciences consistently exhibit the highest share of female authors, maintaining a leading position throughout the century and approaching parity at nearly 50% by 2020. The social sciences and life sciences follow a similar, slightly lower trajectory, reaching approximately 45%. This leading position likely reflects earlier, more comprehensive shifts in gender norms regarding social, medical, and health professions. Correspondingly, the mitigation of the Matilda effect began earlier in these fields. In the health sciences, the ratio began a steady decline in the 1980s, crossing below 1 by the early 2000s. In the social sciences, the ratio remained notably elevated for most of the century but experienced a steep, uninterrupted decline starting in the 1990s, dropping well below the parity line by 2020. While the life sciences successfully crossed below the parity line alongside the other disciplines, the data uniquely shows a slight stabilization in the ratio during the final years of the observation period.

The physical sciences consistently display the lowest overall share of female authors, with the share barely exceeding 30% by the 2020s. This substantial lag possibly highlights entrenched, systemic barriers to entry and retention within math-intensive and engineering disciplines, often attributed to historical stereotyping, a lack of early female role models, and a predominantly male-dominated institutional culture in previous years. Crucially, however, the development of the Matilda effect in the physical sciences mirrors the other, more gender-balanced disciplines. Despite the persistent barrier to entry, the structural disadvantage for women already within the discipline regarding corresponding authorship has declined at a similar pace, crossing below 1 in the early 2000s. This suggests that while it remains disproportionately difficult for women to enter and remain in the physical sciences, those who do are currently assuming corresponding authorship roles at rates commensurate with, or slightly exceeding, their male colleagues.

Figure 5 consolidates the longitudinal trajectories of female authorship across different geographic regions from 1900 to 2025, displaying both the overall share of female authors and the corresponding authorship ratio. To ensure decipherability, the trajectories for all regions except Europe and the Anglo-American block were smoothed due to low total case numbers and high volatility in earlier periods. Whenever an author is affiliated with multiple countries, such as having multiple institutional addresses, all respective regions are considered in the analysis. While all regions demonstrate a clear secular trend toward increased female representation over the observation period, reflecting a general decline in the Matilda effect across all regions, the data reveals a highly stratified global landscape with distinct regional trajectories.

Excluding the Anglo-American block, the Americas (representing Latin America and the Caribbean) consistently lead in female authorship from the late 1980s onward. This region is the only one to closely approach gender parity alongside other regions, reaching approximately 45% by the end of the observation period. A relatively high rate of women's participation in this region has also been reported by Sugimoto et al. (2013) and Holman, Stuart-Fox and Hauser (2018). The regional disparities in female authorship can be contextualized through macro-sociological differences and regional academic labor markets. The leading position of Latin America may be attributed to the region's specific academic labor market dynamics, where academic and research positions are concentrated in the public sector. Because these public academic roles are historically characterized by instability of funding compared to the security offered by the private sector, they may have been less monopolized by men (Pradier et al. 2025). Consequently, academia became a more accessible and feminized sector compared to the Global North. Correspondingly, for all regions except Asia, we see the corresponding authorship curve going below 1, even if only very slightly, indicating a marginal over-representation of women as corresponding authors in recent years. Furthermore, Africa converges with and slightly overtakes the Western baseline in the 21st century.

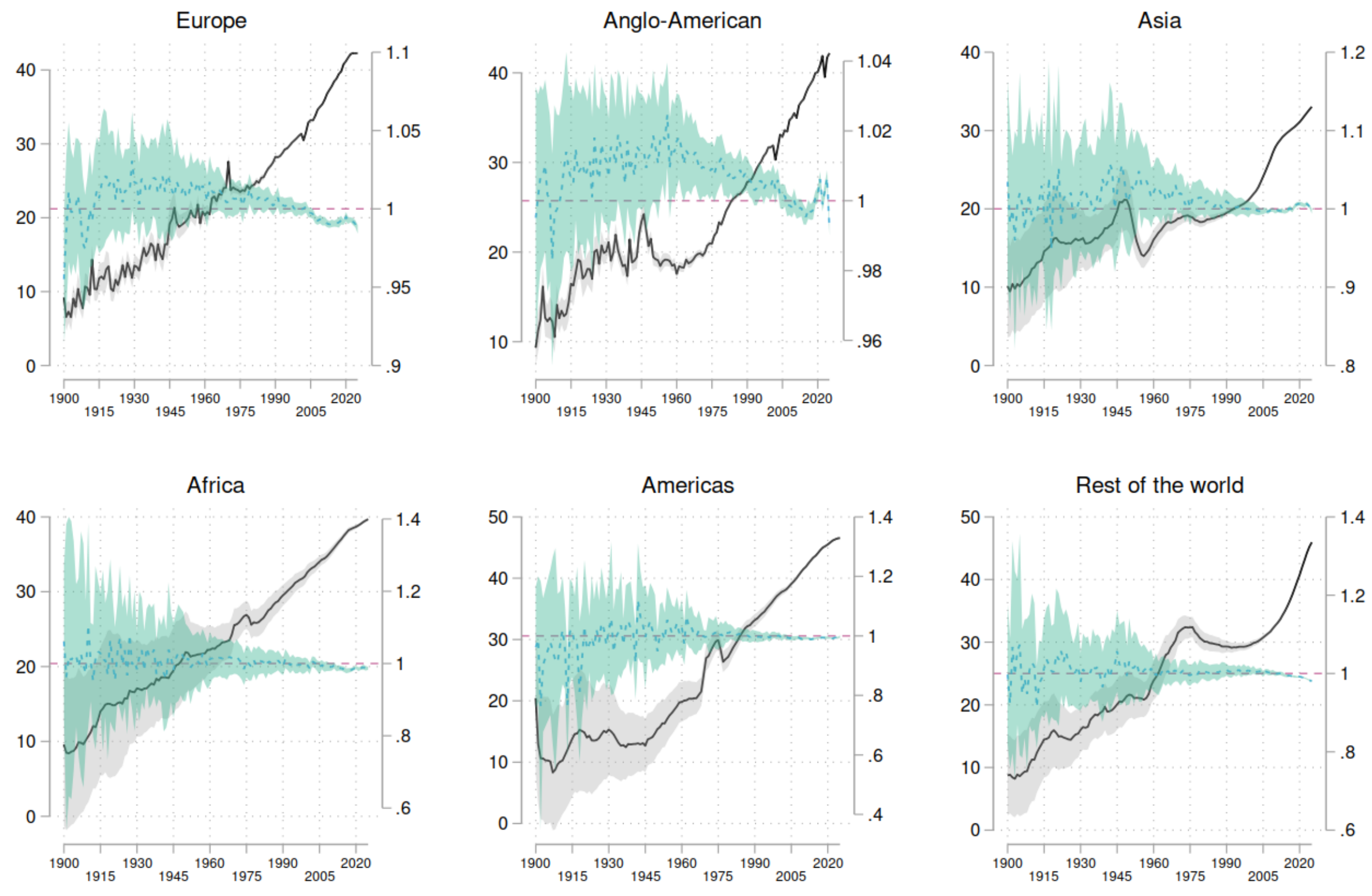

***Figure 5: Longitudinal trends in female authorship and the Matilda effect by geographic region.***
*Notes: The solid line displays the overall share of female authors (left y-axis). The dashed line represents the ratio between the share of female authors and female corresponding authors (right y-axis), serving as an indicator for the Matilda effect. Trajectories for regions other than Europe and the Anglo-American block are smoothed. Europe excludes Russia and Turkey; Anglo-American includes the USA, Canada, Australia, and New Zealand; Asia includes China, South Korea, Japan, and India; Americas includes all Middle and South American countries. Source: OpenAlex, own computations. 95% confidence intervals included.*

The Europe and Anglo-American regions track closely to one another throughout the entire period, representing a steady, institutionalized increase that reaches approximately 40% to 43% by 2025. An outstanding difference, however, is the development directly after the end of World War II. While in Europe, the dip after 1945 remained rather weak, resulting in quite linear growth of corresponding female authors over time, the Anglo-American curve is strongly U-shaped after 1945, indicating a period where women were driven out of the positions they held during the war. A quite similar dip is also present in the Asian countries at that time. Regarding the ratio between female shares and female corresponding authors, it seems that the global reversal in the past 20 years was mainly driven by Europe, as the reversal here is the strongest. For the Anglo-American countries, it seems that the shift to more female leading authors has been reversed again in the last 10 years.

Asia persistently exhibits the lowest share of female authors. Although following the general upward trend, the gap between Asia and the rest of the world has remained substantial, with female representation only reaching about 35% by 2025. Similar results for Asia can also be found in other studies; Bendels et al. (2018) report, for example, “a non-advanced integration of female scientists” for Japan. The persistent lag in the Asian region may largely reflect the data weight of massive scientific producers like Japan, South Korea, and increasingly China. Academic cultures in these countries have historically been characterized by rigid hierarchical structures, intense working-hour expectations, and strong traditional gender norms regarding caregiving (Brinton Mary and Oh 2019). These compounding factors can create ‘leaky pipeline’ effects, hindering female progression and retention in publishing academic roles. In Asia, it seems that the corresponding authorship curve never fell below 1, meaning that male authors are still dominating.

## 3.2 Robustness checks of the findings based on imputed data

To ensure the validity of our primary findings, we conducted a robustness check by repeating the core analyses using raw, unimputed data. While statistical imputation is a widely accepted methodological procedure to account for missing metadata and prevent sample attrition, it relies on underlying assumptions that could potentially mask or artificially distort specific empirical patterns. To rule out this possibility, Figure 6 visualizes our two primary metrics – the overall share of female authors and the corresponding authorship ratio – relying on a complete case analysis. In this restricted dataset, only publications where the gender of all listed authors could be definitively assigned are included. As expected, excluding all imputed records significantly reduces the sample size, which naturally translates to larger statistical volatility and wider confidence intervals, particularly within the historically sparse datasets of the early 20th century. However, despite this increased noise, the fundamental secular trends remain remarkably stable.

The solid line in Figure 6 confirms the long-term upward trajectory in overall female authorship, maintaining the distinct historical shocks discussed previously. Crucially, the analysis of the Matilda effect, represented by the dashed ratio line, mirrors the results of the imputed main model. The ratio remains consistently larger than 1 throughout the 20th century, confirming the persistent historical underrepresentation of women in corresponding authorship roles. Furthermore, the pivotal structural shift is still clearly visible, with the trend reversing and the curve falling below the parity line of 1 just after the turn of the millennium. While the raw data exhibits a slight, singular deviation in the form of increased volatility in the most recent years – a phenomenon likely attributable to standard data collection issues or indexing lags in recent database updates – the overarching long-term trajectory is unaffected. Ultimately, this complete case analysis demonstrates that our core conclusions regarding the historical evolution of the Matilda effect are robust and are not an artifact of the chosen imputation strategy.

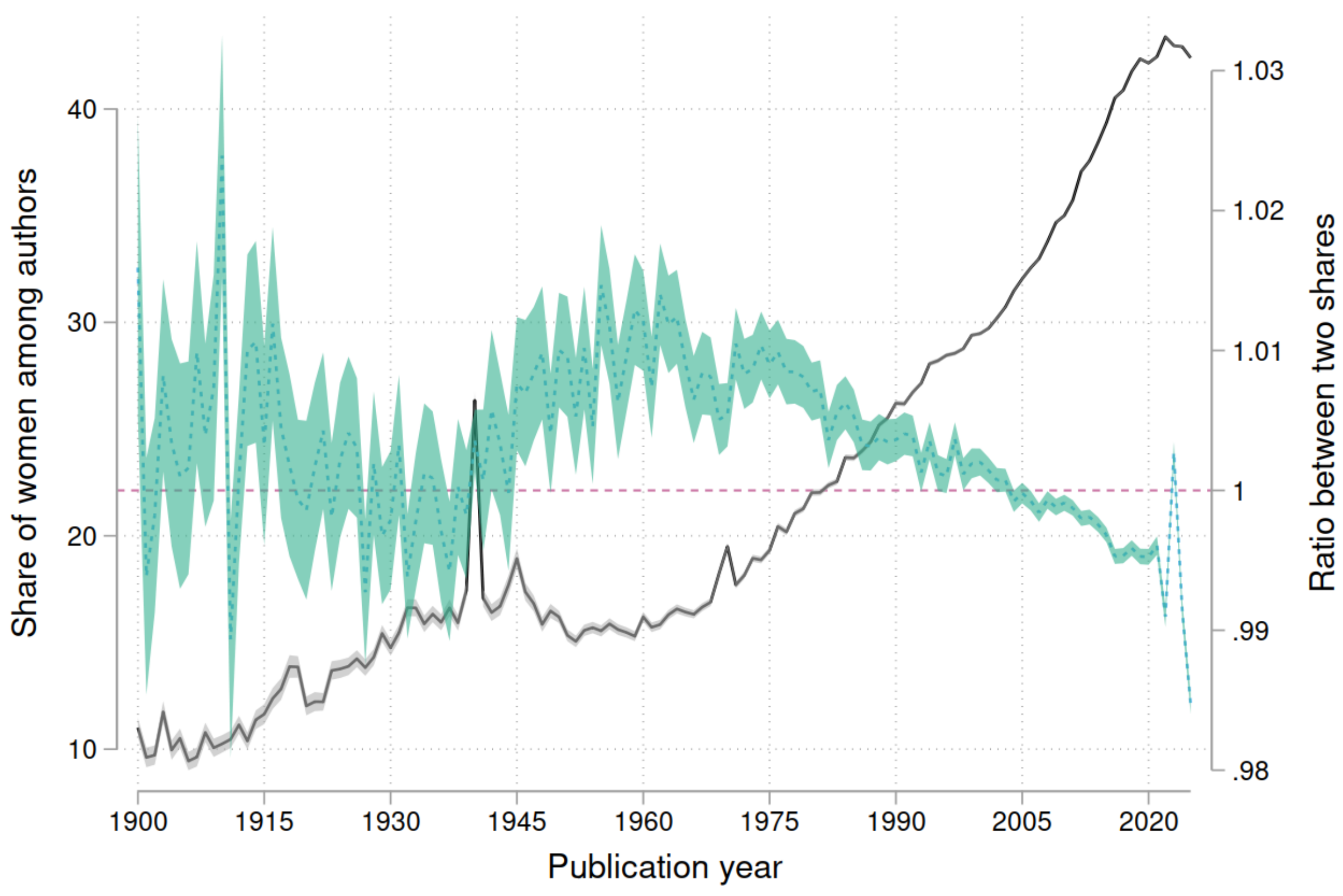

***Figure 6: Robustness check of female authorship and the Matilda effect using unimputed raw data.***
*Notes: Complete case analysis. The solid line displays the overall share of women among all authors (left y-axis). The dashed line represents the ratio between the share of female authors and female corresponding authors (right y-axis). Included are publications where gender information is available for all listed authors (no imputation). Source: OpenAlex, own computations. 95% confidence intervals included.*

As a second robustness check, we repeated the primary analysis by excluding publications with zero citations. The rationale behind this restriction is to focus on research that has impacted the scientific community and is actively used by other researchers in their work, acting as a possible proxy for higher scientific quality. By applying this filter, the total analytical corpus is reduced to 99.9 million publications, reflecting the reality that a substantial portion of academic literature is never cited. For this model, we retained our initial data imputation strategy, as the number of citations is a complete variable without any missing value. As illustrated in Figure 7, despite this sample restriction, the longitudinal patterns remain similar to the main findings. The overall share of female authors continues its steady upward trajectory throughout the observation period.

The evolution of the Matilda effect – represented by the dashed ratio line in Figure 7 – mirrors the primary model. The ratio confirms a persistent structural disadvantage for women in corresponding authorship roles for most of the 20th century, followed by a continuous decline that eventually crosses the parity threshold. However, a distinct deviation occurs in most recent years, where the general trend of the ratio curve tends to increase. Rather than indicating a resurgence of the Matilda effect, this recent uptick is probably an artifact of the sample selection. Because it inherently takes some time until recent publications can even be cited in new works, the subset of recent publications that are already cited is highly selective. Consequently, this right-edge artifact does not undermine the overarching historical trend. This robustness check thus seems to demonstrate that our main results and conclusions hold true even for cited studies, showing the general stability of our findings.

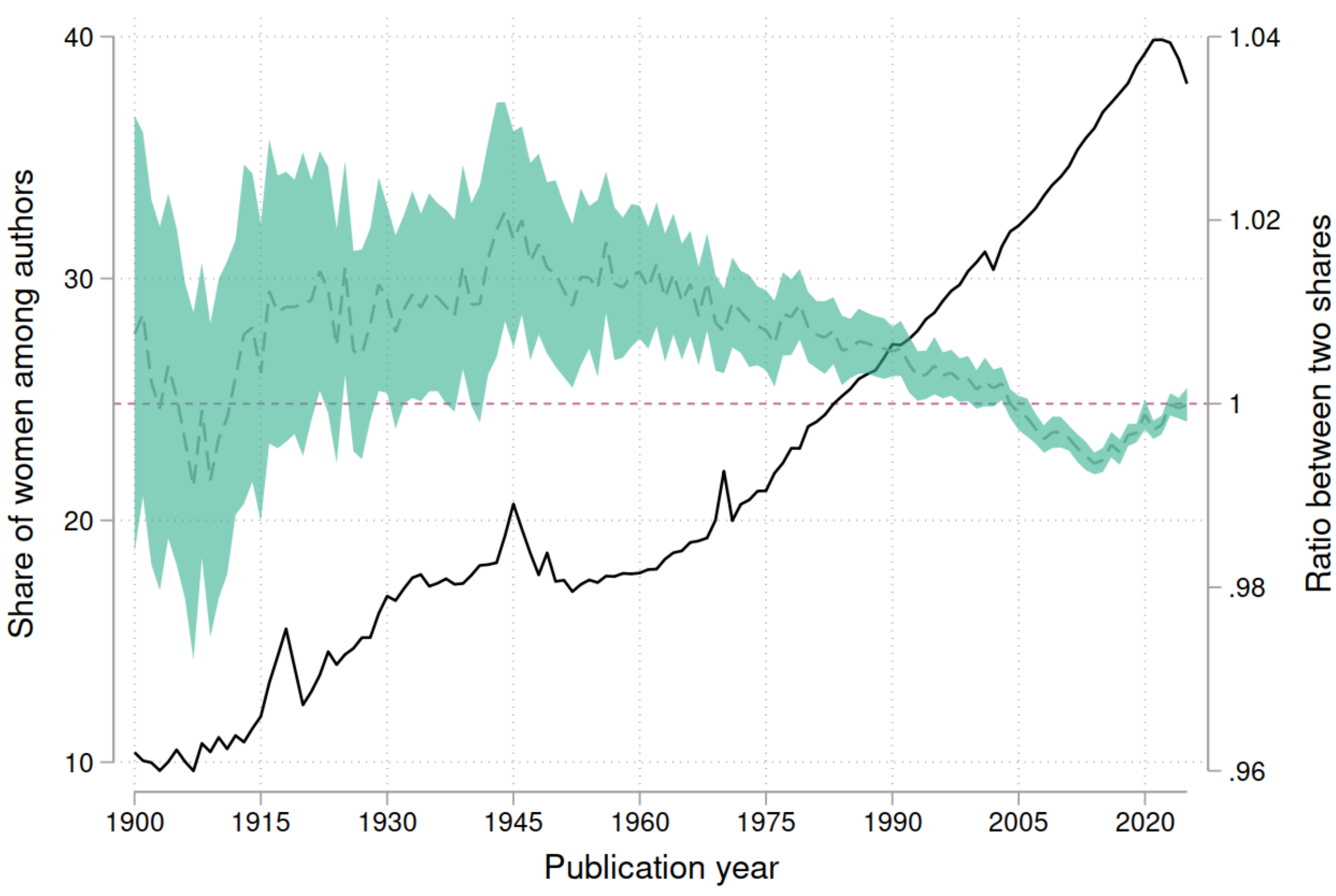

***Figure 7: Robustness check of female authorship and the Matilda effect for cited publications.***
*Notes: The solid line displays the overall share of women among all authors (left y-axis). The dashed line represents the ratio between the share of female authors and female corresponding authors (right y-axis). The sample is restricted to publications that were cited at least once, reducing the corpus to 99.9 million records. Imputed data is used. Source: OpenAlex, own computations. 95% confidence intervals included.*

# 4 Discussion

This study investigated the historical evolution of female participation in science and the prevalence of the Matilda effect over the past 125 years. Using the OpenAlex database, we analyzed over 220 million publications from 1900 to 2025 and inferred the binary gender of approximately 60 million authors using a combination of deterministic approaches and LLMs. For most of the 20th century, the ratio between the shares of female participation and female corresponding authors remained above 1, confirming a historical Matilda effect where women are systematically underrepresented in leadership roles relative to their overall participation. As the ratio is around 1.005 up until the 1990s, there were, on average, 5 more male corresponding authors per 1000 publications than expected, given the share of female authors. This means women were disadvantaged twice: while there were much fewer female scientific authors in general (only around 25% globally up until the 1990s), these fewer authors had even lower chances of being in the lead. However, over the past two decades, this trend has inverted. If a pronounced Matilda effect were currently present, one would expect a substantial divergence between the two trajectories (share of females among authors and share of females among corresponding authors). Instead, the convergence indicates that this specific manifestation of the effect has declined in academic publishing. Nowadays, women are slightly overrepresented among corresponding authors, relative to the overall share of women being authors. However, as the total effect size is very close to one, given our data, it seems that gender influences on lead roles are almost balanced nowadays. This is valid for all disciplines and almost all world regions.

The overall trajectory of female participation shows a continuous increase, rising from approximately 10% in the 1910s to over 40% in recent years. This growth was slow and irregular between 1910 and 1960, but it accelerated significantly thereafter. Structural breaks are visible during global wars, where women temporarily filled research roles left by men, followed by temporary setbacks during post-war demobilization when returning men reclaimed their positions. Despite these disruptions and a recent slowdown in growth rates prior to the global COVID-19 pandemic, the long-term positive trend remains unambiguous. Notable exceptions to these general trends exist across disciplines and geographic regions. The physical sciences persistently exhibit the lowest share of female authors, which may reflect institutional factors, historical stereotyping, or lower average interest. Regionally, several Asian countries (including India, China, Japan, and South Korea) demonstrate female participation rates approximately ten percentage points lower than other regions, likely reflecting distinct cultural traditions and academic environments. The broader literature provides context for these observations, highlighting that structural vulnerabilities, such as caregiving obligations and 'sticky steps' in career progression, continue to drive women out of the academic pipeline and reduce career longevity (Fusulier, Barbier and Dubois-Shaik 2017; Meirmans, Lamatsch and Neiman 2022).

Several limitations must be considered when interpreting these findings. First, while OpenAlex provides comprehensive coverage, no database is entirely complete, and the data contains minor artifacts, such as artificial spikes in the total number of publications at decade intervals. Second, gender classification is limited by missing information, particularly when only author initials are available. Although we utilized a two-step approach combining deterministic libraries with LLMs, the latter operate as black boxes, making the exact reasoning behind specific classifications opaque. Despite these classification challenges, recent research affirms the general utility of LLMs for this purpose (Goyanes et al. 2025), indicating that our research benefits from the approach even if misclassifications cannot be completely ruled out. Third, a purely bibliometric perspective offers a constrained view of a highly complex reality. Quantitative indicators capture only a fraction of total research output, especially in disciplines like the qualitative social sciences, humanities, and engineering, where journal articles are not the exclusive carrier of scientific knowledge (Mongeon and Paul-Hus 2016; van den Besselaar and Sandström 2020). Fourth, publications vary profoundly in their intrinsic quality and scientific significance, meaning that simple counts do not fully represent research weight.

## Acknowledgements

During the preparation of this work the authors used Gemini Pro 3.1 to improve language and readability. After using this tool, the authors reviewed and edited the content as needed and take full responsibility for the content of the publication.